\documentstyle[amssymb,aps,epsf,prl,multicol]{revtex}

\begin{document}
\title{{\large Rotational Cooling of Polar Molecules by Stark-tuned Cavity Resonance%
}}
\author{C. H. Raymond Ooi\thanks{{\footnotesize ooi@spock.physik.uni-konstanz.de}}}
\author{{\it Fachbereich Physik der Universit\"{a}t Konstanz, Fach M674}, D-78457}
\address{{\it Konstanz, Germany}}
\date{\today}
\maketitle

\begin{abstract}
A general scheme for rotational cooling of diatomic heteronuclear molecules
is proposed. It uses a superconducting microwave cavity to enhance the
spontaneous decay via Purcell effect. Rotational cooling can be induced by
sequentially tuning each rotational transition to cavity resonance, starting
from the highest transition level to the lowest using an electric field.
Electrostatic multipoles can be used to provide large confinement volume
with essentially homogeneous background electric field.
\end{abstract}

\section{Introduction}

Currently, there are three main techniques to produce translationally cold
molecules. The buffer gas \cite{buffer} and Stark deceleration \cite{Meijer}
cooling schemes employ trapping of the low field seekers which requires the
molecules to be in the appropriate internal states. Cold molecules produced
from photoassociation technique \cite{PA} are usually vibrationally hot,
occupying a number of high lying vibrational states. Vibrational cooling
schemes using optimal control of ultrashort pulses have been proposed \cite
{vibcool}. Translationally cold molecules are useful for precision
spectroscopy and measurements, molecular optics and interferometry as well
as cold collisions studies, although they may be internally hot. It is also
desirable to have translationally and internally cold molecules as well if
the aim is to obtain molecular Bose-Einstein Condensate \cite{molecular BEC}%
. For the purpose of trapping, the molecules need to be cooled internally.

There has been no scheme to rotationally cool molecules although there are
several methods of producing molecules in a single rotational state. A
classic example is the electrostatic low-J selector which has been used for
rotational state selection long ago \cite{low J selector}. Optical
Stern-Gerlach has been demonstrated for atoms \cite{OSG} and molecular state
selection by lasers has been proposed \cite{Domokos}. However, these
techniques are best applied to molecular beam. They are not cooling schemes
since they do not employ a dissipative mechanism. Supersonic expansion is
based on dissipation and may also produce rotationally cold molecules, but
cannot be used to cool molecules which are already translationally cold
since it requires high pressure and can only be applied mechanically on hot
molecules. Besides, for translationally cold molecules it is more efficient
to employ dissipative process for internal cooling which avoids the loss of
molecules.

Recently, we have proposed 1-D translational cooling schemes for molecules
which rely on a single optical spontaneous emission \cite{Raymond}. The
schemes can be repeated for 3-D cooling if the 1-D cooled molecules can be
brought back to the initial internal state again. Molecules in excited
rot-vibrational states can redistribute significant amount of the internal
energy into the translation motion by state-changing inelastic collisions.
At high temperature, this becomes a problem for trapping due to the
increased collision rate. Therefore, it is essential to removed the internal
excitations by spontaneous emissions fast enough before the inelastic
collisions occur. Spontaneous emission carries away entropy to radiation
while inelastic collision transfers entropy to the translational degree. For
sufficiently dilute heteronuclear molecules, vibrational cooling occur
within $0.1s-1s$ timescale through infrared spontaneous emissions before the
vibrational inelastic collisions take place. Thus, the vibrational entropy
is discarded as radiation entropy instead of the translational entropy.
However, rotational spontaneous emissions can occur only in polar molecules
and take much longer time, beyond the experimental timescale for dilute gas.
For dense gas, rotational decay may be enhanced through many-body effect,
but inelastic collision rate may become dominantly large.

In this paper, we propose a rotational cooling scheme for confined polar
molecules. We use external electric field to tune the internal transitions
into resonance with a lossy microwave cavity (Fig. \ref{rotcavEtune}). This
enhances the rotational spontaneous emissions via the Purcell effect \cite
{Purcell} whereby the populations are transferred to a lower rotational
level in stages and eventually to the ground rotational level. The Purcell
effect has been experimentally demonstrated by Goy et. al. using Rydberg
atoms in a microwave cavity \cite{Goy}. The cooling process requires long
interaction time and we have proposed a method to confine the molecules
within the cavity field.

\section{Initial State}

We consider a gas of polar diatomic molecules at thermal equilibrium
temperature $T$ \ with the Maxwell-Boltzmann distribution. We also assume
that the gas is sufficiently dilute such that the inelastic collision is
negligible throughout the cooling process. At below room temperature, the
populations are essentially in the eigenenergy in the ground electronic
state $|g,\Omega \rangle .$ Thus, the population and the eigenenergy in the
vibrational state $|n\rangle $ and rotational states $|J,M\rangle $ with $%
J\geq \Omega $ (the total electronic angular momentum) are given
respectively by

\begin{eqnarray}
P(g,\Omega ,n,J) &=&P_{o}(2J+1)\sum\limits_{\Omega =\frac{1}{2},\frac{3}{2}%
}\exp \{-E_{g,\Omega ,nJ}/k_{B}T\}  \label{initial pop} \\
E_{g,nJ,\Omega } &=&hc\{T_{g}+\omega _{g}(n+\frac{1}{2})-\omega _{g}x_{g}(n+%
\frac{1}{2})^{2}+(B_{g}-\alpha _{g}(n+\frac{1}{2}))(J(J+1)-\Omega ^{2})\}
\label{E}
\end{eqnarray}
where $T_{g},\omega _{g},\omega _{g}x_{g},$ $B_{g}$ and $\alpha _{g}$ are
the standard electronic, vibrational and rotational constants (in
wavenumber) \cite{Herzberg}.

\vspace{0in}From Eqs. \ref{initial pop} and \ref{E}, we estimate that for
CsF molecule with ($\omega _{e},\omega _{e}x_{e},B_{e},\alpha _{e}$)=($%
352.56,1.61,0.184,0.0012$)cm$^{-1}$ \cite{Herzberg}, only the first 5
rotational levels in the ground vibrational level ($n=0$) are significantly
occupied at temperature around $1K$. The rotational levels and the magnetic
(Zeeman) states for cases $\Omega =0$ and $\Omega =\frac{1}{2}$ are shown in
Fig. \ref{EJCsF}a and Fig. \ref{EJOH}a respectively. Initially, the
molecules are in all the magnetic states. In order to cool the molecules in
all the magnetic states, the cavity must be able to support linearly
polarized as well as circularly polarized photons corresponding to $\Delta
M=0$ and $\pm 1$ transitions respectively. If we assume that the cavity
supports only linearly polarized photons, only the states with $M\leq \pm
\Omega $ will be sequentially Stark-tuned to cavity frequency by an external
static electric field for enhanced spontaneous\ decay toward the ground
level. Molecules in other states (with $M>\pm \Omega $) are off-resonance
with the cavity and their decays will not be enhanced. In this case, we will
have to extract only the molecules in the states $M\leq \pm \Omega $ for
rotational cooling.

\section{Rotational Cooling Scheme}

In this section, we elaborate on the rotational cooling scheme based on the
proposed setup apparatus (Figure ~\ref{rotcavEtune}).

\subsection{Cooling mechanism}

The spacings of rotational levels in polar diatomic molecules increase
monotonically with energy levels. We use this monotonic property to bring
the populations from the high $J$ levels to a low $J$ (ground) level by
enhanced rotational decay in a {\em lossy} microwave cavity. We want to
bring each pair of transition ($|J,M\rangle $ to $|J-1,M-q\rangle $ with $%
q=0,\pm 1$ and $J\geqslant 1$) into resonance with the cavity at each time,
starting from the $J_{\max }\rightarrow $ $J_{\max }-1$ transition and
ending with the $J=\Omega +1$ $\rightarrow $ $J=\Omega $ transition. We
shall elaborate on the simplest scheme employing only the $\pi $-transitions
($q=\Delta M=0$) (Figs. \ref{EJCsF}a and \ref{EJOH}a), as well as the scheme
which uses multiple polarized cavity for both the $\sigma $- dan $\pi $%
-transitions.

\subsection{Cavity Enhanced Decay}

When the cavity is resonant with the radiative transition from $|J,M\rangle $
to $|J-1,M-q\rangle ,$ the decay rate is enhanced, given by the Purcell
formula \cite{Scully QO}
\begin{equation}
\Gamma _{c,J,M,q}=\frac{2\mu _{J,M,q}^{2}}{\varepsilon _{o}V\hbar }Q=\eta
\Gamma _{o,J,M,q}  \label{Purcel formula}
\end{equation}
where $\Gamma _{o,J,M,q}=\frac{16\pi ^{3}\mu _{J,M,q}^{2}}{3\varepsilon
_{o}h\lambda _{c}{}^{3}}$ is the free space decay rate, $\eta =\frac{%
3\lambda _{c}^{3}Q}{4\pi ^{2}V}$ the cavity enhancement factor, $\mu
_{J,M,q} $ the dipole transition matrix element, $\lambda _{J}$ the
transition wavelength resonant with the cavity frequency $\omega _{c}=2\pi
c/\lambda _{c}$, $Q$ the quality factor and $V=\int |u_{mn}({\bf r}%
)|^{2}d^{3}r$ the effective cavity volume for the Hermite-Gaussian (m,n)
transverse mode function $u_{mn}({\bf r}).$

We consider a Gaussian mode in a symmetric {\em confocal} resonator (aligned
along z-axis, see Fig. \ref{rotcavEtune}), we have \cite{Yariv Qtm Elect}
the cavity spacing $L=(s+\frac{1}{2})\lambda _{c}/2$ for longitudinal mode
order $s$, field waist $w_{o}=\sqrt{\frac{\lambda _{c}L}{2\pi }}=\lambda _{c}%
\sqrt{(s+\frac{1}{2})/4\pi }$ intensity $|u_{00}({\bf r})|^{2}=\frac{%
w_{o}^{2}}{w^{2}(z)}e^{-2(x^{2}+y^{2})/w^{2}(z)}$ with $%
w^{2}(z)=w_{o}^{2}(1+(\frac{2\lambda _{c}z}{\pi w_{o}})^{2}).$ From these,
the effective volume and the enhancement factor is rewritten as

\begin{eqnarray}
V &=&Lw_{o}^{2}\pi /2=\frac{\lambda _{c}L^{2}}{4}=(s+\frac{1}{2})^{2}\lambda
_{c}^{3}/16  \label{effective volume} \\
\eta &=&\frac{12Q}{\pi ^{2}(s+\frac{1}{2})^{2}}  \label{enhancement factor}
\end{eqnarray}

Large enhancement factor is obtained by using a superconducting cavity with
large $Q$ and a low order mode $s.$ The free and enhanced decay rates are
dependent on the $J$ and $M$ through the numerical factor in the matrix
elements.

\subsection{Shielding thermal photons}

In microwave regime, the mean number of thermal photons $\bar{n}(\omega
_{c}) $ is not negligible. It enhances the spontaneous decay rate as $\Gamma
_{o,J,M,q}=\frac{\mu _{J,M,q}^{2}\omega _{c}{}^{3}}{3\varepsilon _{o}\hbar
\pi c^{3}}(2\bar{n}(\omega _{c})+1).$ However, it reduces the cooling
efficiency since the decaying states cannot be completely depopulated due to
simultaneous incoherent excitations of the thermal photons. The usual way to
reduce the thermal photons is by cooling of the apparatus. The inner surface
of the apparatus is coated with microwave photon absorbing material like
graphite while the outer surface can also be shielded with a photonic
band-gap structure \cite{PBG}. Thus, we can set $\bar{n}=0$ and write the
free space decay rate for $\pi $-transition $|J,M\rangle \rightarrow
|J-1,M\rangle $ and $\sigma $-transition $|J,M\rangle \rightarrow |J-1,M\mp
1\rangle $\cite{Kroto} respectively as

\begin{equation}
\Gamma _{o,J,M,q}=\frac{||\mu ||^{2}}{\lambda _{c}{}^{3}}g(J,M)\allowbreak
2.\,\allowbreak 818\,7\times 10^{46}\text{( in S.I.)}  \label{Gamm0}
\end{equation}
where $||\mu ||$ is the reduced electric dipole matrix element for the
vibronic state, and $g(J,M)=\frac{(J+M)(J-M)}{(2J-1)(2J+1)}$ and $\frac{%
(J\pm M+1)(J\pm M+2)}{(2J-1)(2J+1)}$ for the $\pi $-transition and $\sigma $%
-transition respectively.

\subsection{Electric field tuning}

In order to enhance the decay of each transition, we can either tune the
cavity or the levels into resonance with each other. In principle, we can
construct a tunable cavity by changing the distance between the mirrors.
First, the molecules are put into the cavity tuned to enhance the
spontaneous emission of the highest rotational state, from $J_{\max }$ to $%
J_{\max }-1$. Then, the cavity is tuned to the next lower transition and so
on until all the populations are brought to a single rovibrational state $%
J=0.$ Here, $L$ changes from $(s+\frac{1}{4})\lambda _{J_{\max }}/2$ to $(s+%
\frac{1}{4})\lambda _{\Omega }/2.$ However, this approach is not very
feasible in practice \cite{Pepijn Pinkse}.

It is more realistic to tune the transition levels into resonance with
cavity frequency using an {\em external electric field}. For $\Delta M=0$
transitions, photons are most likely to be emitted perpendicular to the
dipole. To maximize the molecular dipole-cavity field coupling strength.and
the spontaneous decay rate, the dipole should be parallel to the {\em cavity
electric field} (in x-y plane). Therefore, the tuning electric field is
applied along the {\em x-axis} so that it aligns the dipole along the cavity
field (see Fig. \ref{rotcavEtune}). Here, the $\Delta M=\pm 1$ transitions
can be enhanced too as photons are emitted in all directions, although the
most probable direction is parallel to the dipole. If the tuning electric
field is along the {\em cavity axis}, the $\Delta M=0$ transition may not be
optimumly enhanced because no photon is emitted exactly along the dipole.
However, there is still some enhancement from off-axis modes because the
microwave cavity is essentially a closed cavity which provides a large solid
angle of mode confinement.

A pair of closely spaced electrostatic plates can be introduced into the
cavity to produce a tunable homogeneous transverse electric field $E_{J}$
perpendicular to the cavity axis (say along x-direction) for Stark tuning.
Thus, the edges of the superconducting cavity mirrors can be shielded from
the strong electric field using a metallic casing or coating to maintain a
constant high $Q$ value.

The energy separation between states $|J,M\rangle $ and $|J-1,M-q\rangle $
in the electric field $E_{J,M,q}$ is

\begin{equation}
\Delta {\cal E}_{n,J,M,q}^{n}=hc2J\{B_{e}-\alpha _{e}(n+\frac{1}{2})\}+\frac{%
(\mu E_{J,M,q})^{2}}{2hcB_{e}}(f_{J,M}\allowbreak -f_{J-1,M-q}\allowbreak )
\label{level spacing}
\end{equation}
where $f_{J,M}\allowbreak -f_{J-1,M-q}\allowbreak =\allowbreak \frac{%
J(J+1)-3M^{2}}{J\left( J+1\right) \left( 2J-1\right) \left( 2J+3\right) }%
-\allowbreak \frac{J(J-1)-3(M-q)^{2}}{(J-1)J\left( 2J-3\right) \left(
2J+1\right) }$ \ and $\mu $ is the electric dipole moment.

$\allowbreak \vspace{0in}\allowbreak \allowbreak $From Eq. \ref{level
spacing}, it is possible to keep a pair of transition at cavity resonance $%
\Delta {\cal E}_{n,J,M,q}^{n}=\frac{hc}{\lambda _{c}}$ by varying the
electric field for each set of $J,M,q$. Setting $n=0,$ the required electric
field for each transition can be predicted as

\begin{equation}
E_{J,M,q}=\frac{hc}{\mu }\sqrt{\frac{2B_{e}}{\allowbreak f_{J,M}\allowbreak
-f_{J-1,M-q}\allowbreak }\{\frac{1}{\lambda _{c}}-J(2B_{e}-\alpha _{e})\}}
\label{E tune}
\end{equation}

\begin{figure}[t]
\center \epsfxsize=7cm \epsffile{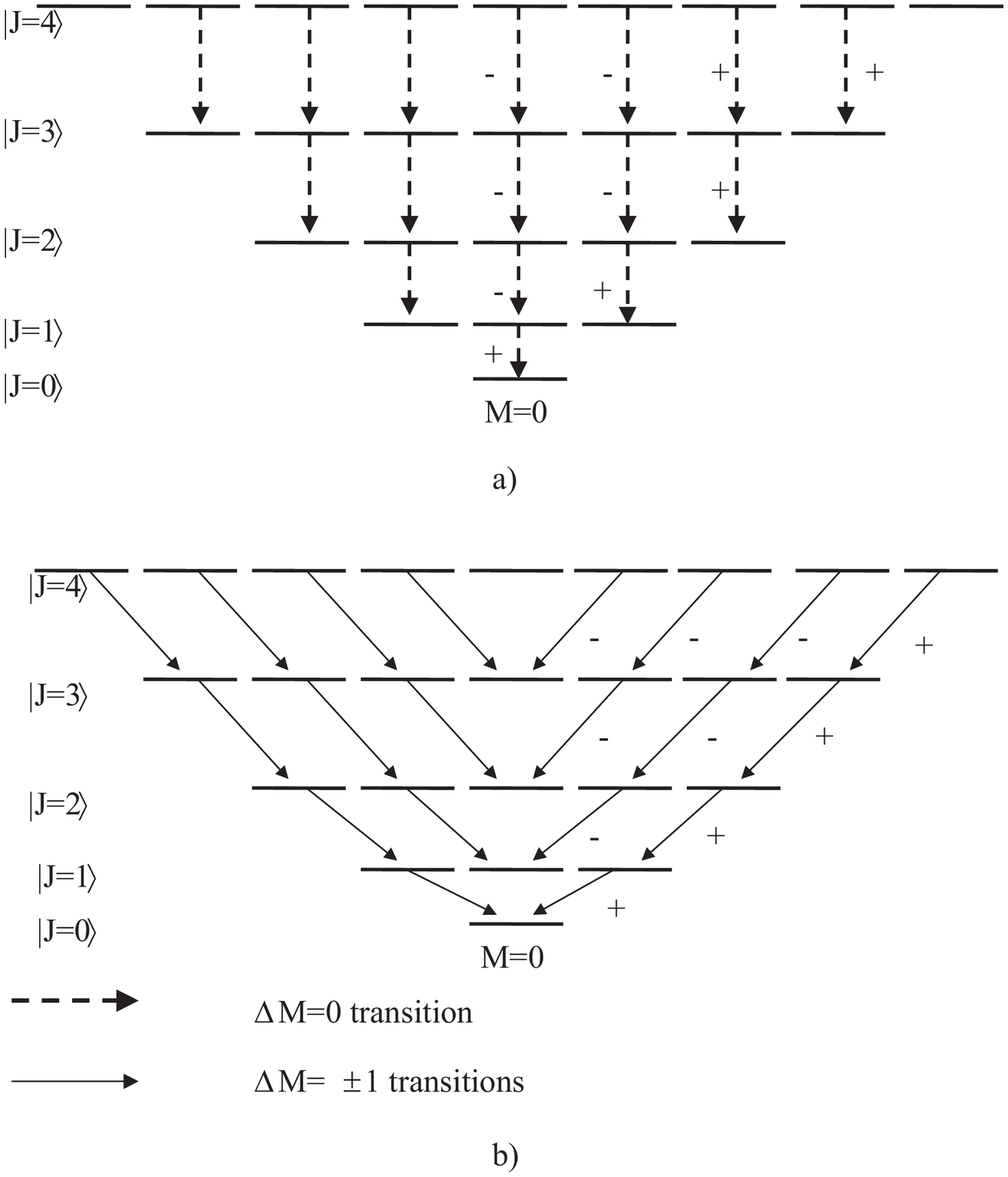}\caption{The signs of
$f_{J,M^{\prime }}\allowbreak -f_{J-1,M}$ in Eq. \ref {E tune}
for: a)$\protect\pi $-transitions ($\Delta M=0$) and
b)$\protect\pi $-transitions ($\Delta M=\pm 1$). Due to the
quadratic dependence of the Stark shift on $M$, the transitions
$|J,|M|+1\rangle \rightarrow |J-1,|M|\rangle $ are symmetrical to
$|J,-|M|-1\rangle \rightarrow |J-1,-|M|\rangle .$} \label{ffsign}
\end{figure}

We need to fix the cavity parameters through $\lambda _{c}$ such that the
argument in the square root is positive. In general, $f_{J,M}\allowbreak
-f_{J-1,M-q}$ can be positive or negative depending on the transition, so it
may not be possible to use a single value of $\lambda _{c}$ to tune all
transitions to cavity resonance. It is only possible to tune those
transitions with negative values of $f_{J,M}\allowbreak -f_{J-1,M-q}$ (the
'negative signs' transitions in Figs. \ref{ffsign}a and b) to cavity
resonance by choosing the cavity parameter $\lambda _{c}$ which gives a
negative nominator in Eq. \ref{E tune}, $\frac{1}{\lambda _{c}}%
<J(2B_{e}-\alpha _{e})$ or $L>\frac{s+\frac{1}{2}}{2J(2B_{e}-\alpha _{e})}$
\ in order to obtain real values of $E_{J,M,q}$. This can be satisfied for $%
J\geqslant 1$ and $J>\sqrt{\frac{1}{2}\left( 1+6M^{2}+\sqrt{\left(
1+3M^{2}+36M^{4}\right) }\right) }$ if we set

\begin{equation}
\frac{1}{\lambda _{c}}=(2B_{e}-\alpha _{e})\text{ or }L=\frac{s+\frac{1}{2}}{%
2(2B_{e}-\alpha _{e})}  \label{L}
\end{equation}
and Eqs. \ref{Gamm0} and \ref{E tune} become

\begin{eqnarray}
\Gamma _{o,J,M,q} &=&||\mu ||^{2}\allowbreak (2B_{e}-\alpha
_{e})^{3}\allowbreak g(J,M)2.\,\allowbreak 818\,7\times 10^{46}\text{( in
S.I.)}  \label{Gamm0 final} \\
E_{J,M,q} &=&\frac{hc}{\mu }\sqrt{\frac{2B_{e}(2B_{e}-\alpha _{e})}{%
\allowbreak f_{J,M}\allowbreak -f_{J-1,M-q}}(1-J)}  \label{E tune final}
\end{eqnarray}
$\allowbreak $

Condition Eq. \ref{L} applies to the $\pi $-transitions (Fig. \ref{ffsign}a)
and the $\sigma $-transitions (Fig. \ref{ffsign}b) where $\allowbreak
f_{J,M}\allowbreak -f_{J-1,M-q}$ are negative. It also applies to $%
|J=1,M=0\rangle \rightarrow |0,0\rangle $ although $\allowbreak
f_{1,0}\allowbreak -f_{0,0}$ is positive (Fig. \ref{ffsign}a) because Eq.
\ref{E tune final} vanishes for $J=1$.

On the other hand, the ''corner'' $\pi $-transitions $|J,M=\pm J\rangle
\rightarrow |J-1,\pm (J-1)\rangle $ have positive denominator $f_{J,\pm
J}\allowbreak -f_{J-1,\pm (J-1)}=\allowbreak \allowbreak \frac{4J+3}{\left(
2J+1\right) \left( 2J+3\right) J\left( J+1\right) }$. The cavity value $%
\lambda _{c}$ defined by Eq. \ref{L} gives negative nominator and cannot be
used to tune these transitions. The ''corner'' transitions and those $\pi $%
-transitions (Fig. \ref{ffsign}a) with 'positive sign' can be tuned for
cooling by using a different cavity parameter which satisfies a positive
nominator in Eq. \ref{E tune}, $\frac{1}{\lambda _{c}}>J(2B_{e}-\alpha _{e})$%
. A reasonable choice is
\begin{equation}
\frac{1}{\lambda _{c}}=J_{\max }(2B_{e}-\alpha _{e})  \label{L positive}
\end{equation}
and we have
\begin{eqnarray}
\Gamma _{o,J,M,q} &=&J_{\max }^{3}||\mu ||^{2}\allowbreak (2B_{e}-\alpha
_{e})^{3}g(J,M)\allowbreak 2.\,\allowbreak 818\,7\times 10^{46}\text{( in
S.I.)}  \label{Gamma0 positive} \\
E_{J,M,q} &=&\frac{hc}{\mu }\sqrt{\frac{2B_{e}(2B_{e}-\alpha _{e})}{%
\allowbreak f_{J,M}\allowbreak -f_{J-1,M-q}}\{J_{\max }-J\}}
\label{E tune positive}
\end{eqnarray}

\begin{figure}[t]
\center \epsfxsize=7cm \epsffile{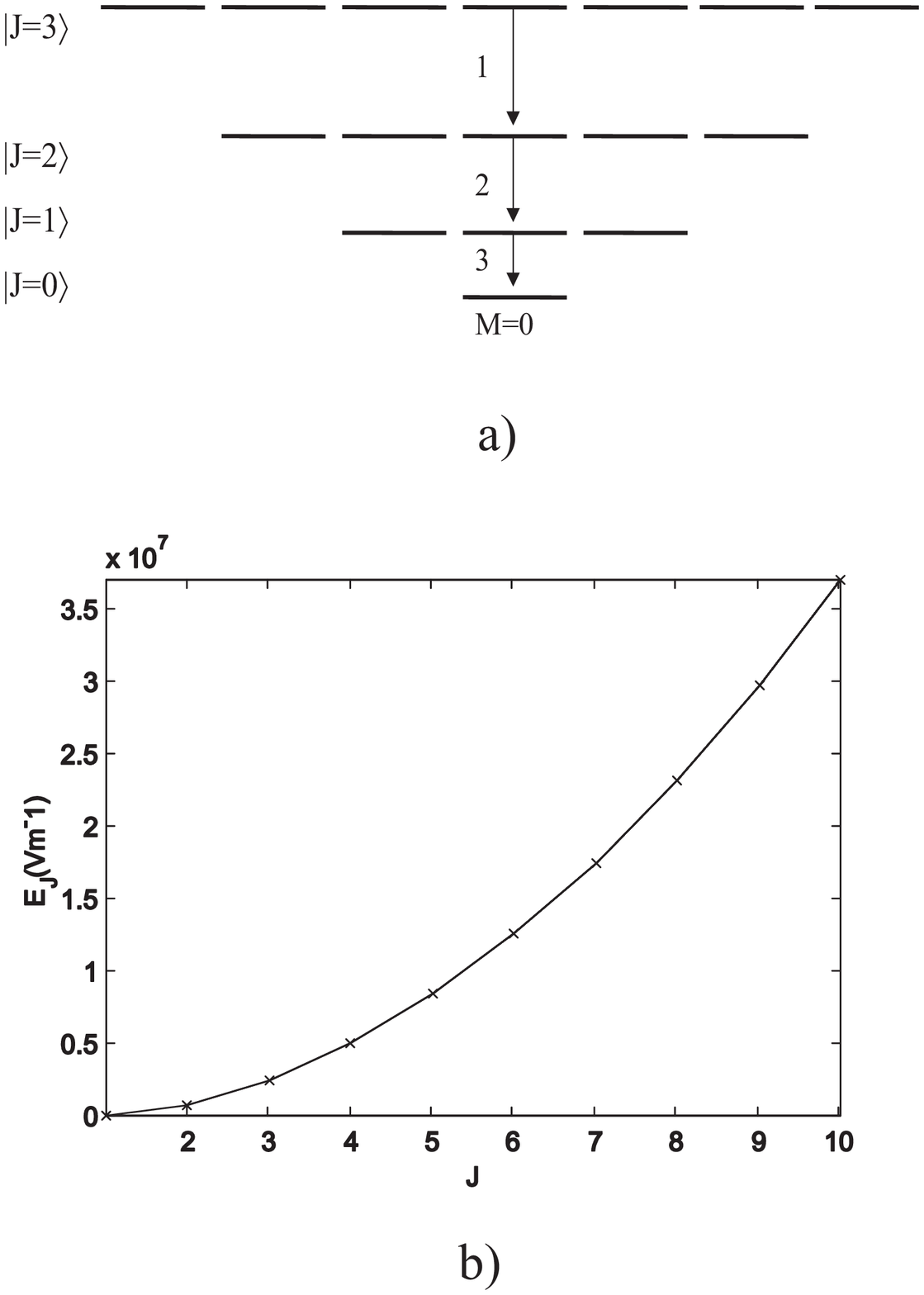}\caption{Spinless (S=0)
molecule like CsF in state $|X^{1}\Sigma _{0},n=0\rangle $ :
a)Rotational levels and the Zeeman states with a single ground
state. Only the parallel transitions 1 to 3 with $\Delta M=0$ are
shown. b) Electric fields for tuning the CsF molecules from the upper state $%
|J_{\max }=10,M=0\rangle $ to the lowest state $|J=0,M^{\prime}=0\rangle $ }
\label{EJCsF}
\end{figure}

\begin{figure}[t]
\center \epsfxsize=7cm \epsffile{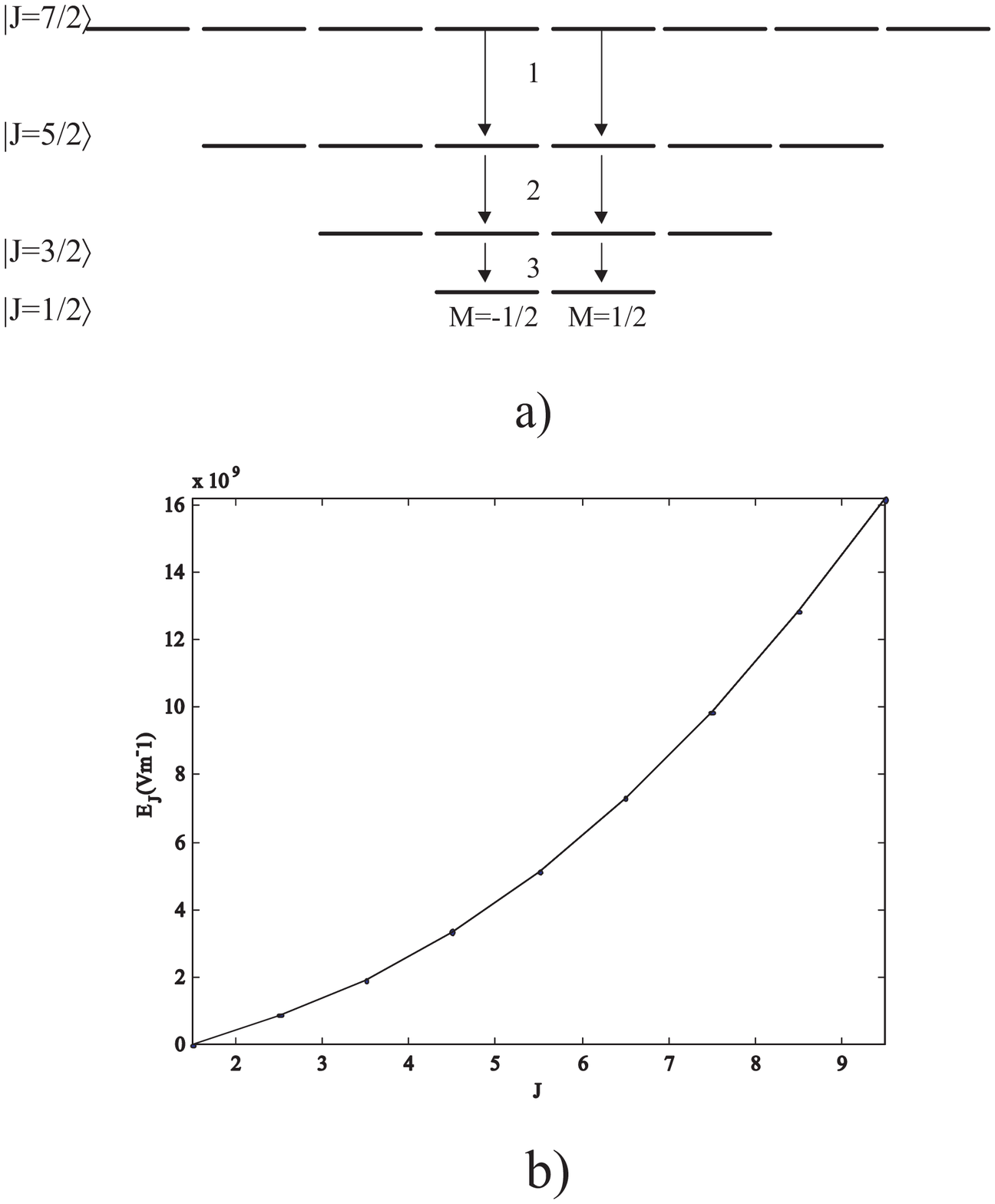}\caption{OH molecule in
state $|X^{1}\Pi_{1/2},n=0\rangle $ : a)Rotational levels and the
Zeeman states with two ground states. Six parallel transitions 1
to 3 are shown. b) Electric fields for tuning the OH molecules
from the upper states of $|J_{\max }=9.5,M=\pm 0.5\rangle $ to
ground states $|J=0.5,M^{\prime}=\pm 0.5\rangle $ } \label{EJOH}
\end{figure}

If the cavity only supports linearly polarized photons, only the parallel or
$\pi $- transitions ($\Delta M=0$) can be enhanced. For the moment, we
consider only the $\pi $-transitions, from the states $|J,|M|\leq \Omega
\rangle $ to $|J-1,|M|\leq \Omega \rangle $. The transitions are tuned to
cavity resonance at each time using the electric field, while all other
Zeeman states are not in resonance with the cavity due to the different
Stark shifts. Only those molecules in the states $|J,-\Omega \leq M\leq
\Omega \rangle $ are rotationally cooled down to a number of $2\Omega +1$
states in the ground level, $|\Omega ,-\Omega \leq M\leq \Omega \rangle $.
For example, with OH molecules in state $^{2}\Pi _{\Omega =\frac{1}{2}}$
(Fig. \ref{EJOH}a), rotational cooling leads to 2 internal ground states.
For spinless molecules like CsF in the ground electronic state $X^{1}\Sigma
_{\Omega =0}$ (Fig. \ref{EJCsF}a) the molecules can be cooled to a single
internal state, $|0,0\rangle .$ The number of tuning steps required for $%
\Delta M=0$ transitions is $(\Omega +1)(J_{\max }-\Omega )$ for integer $%
\Omega $, and $(\Omega +\frac{1}{2})(J_{\max }-\Omega )$ for half-integer $%
\Omega $. The tuning electric fields calculated from Eq. \ref{E tune final}
are shown in Figs. \ref{EJCsF}b and \ref{EJOH}b for different $J$ of the
decaying state.

From Figs. \ref{ffsign}a and \ref{ffsign}b, we see that it is possible
establish a scheme to cool the molecules in most states by using a multiple
field polarizations cavity which can support all the three polarized photons
(linear, $\sigma ^{-}$ and $\sigma ^{+}$ ) corresponding to $\Delta M=0,\pm 1
$ transitions respectively. This enables the molecules in all the Zeeman
states except the 'corner states' $|J,\pm J\rangle ,$ to be Stark-tuned for
cooling into a single ground {\em level }(or single internal {\em state} for
case $\Omega =0$). Due to the quadratic dependence of the Stark shift on $M,$
two pairs of states can be Stark-tuned to cavity resonance at each time; but
only one pair of states for $M=0\Leftrightarrow M=0$ transition.

\begin{figure}[t]
\center \epsfxsize=7cm \epsffile{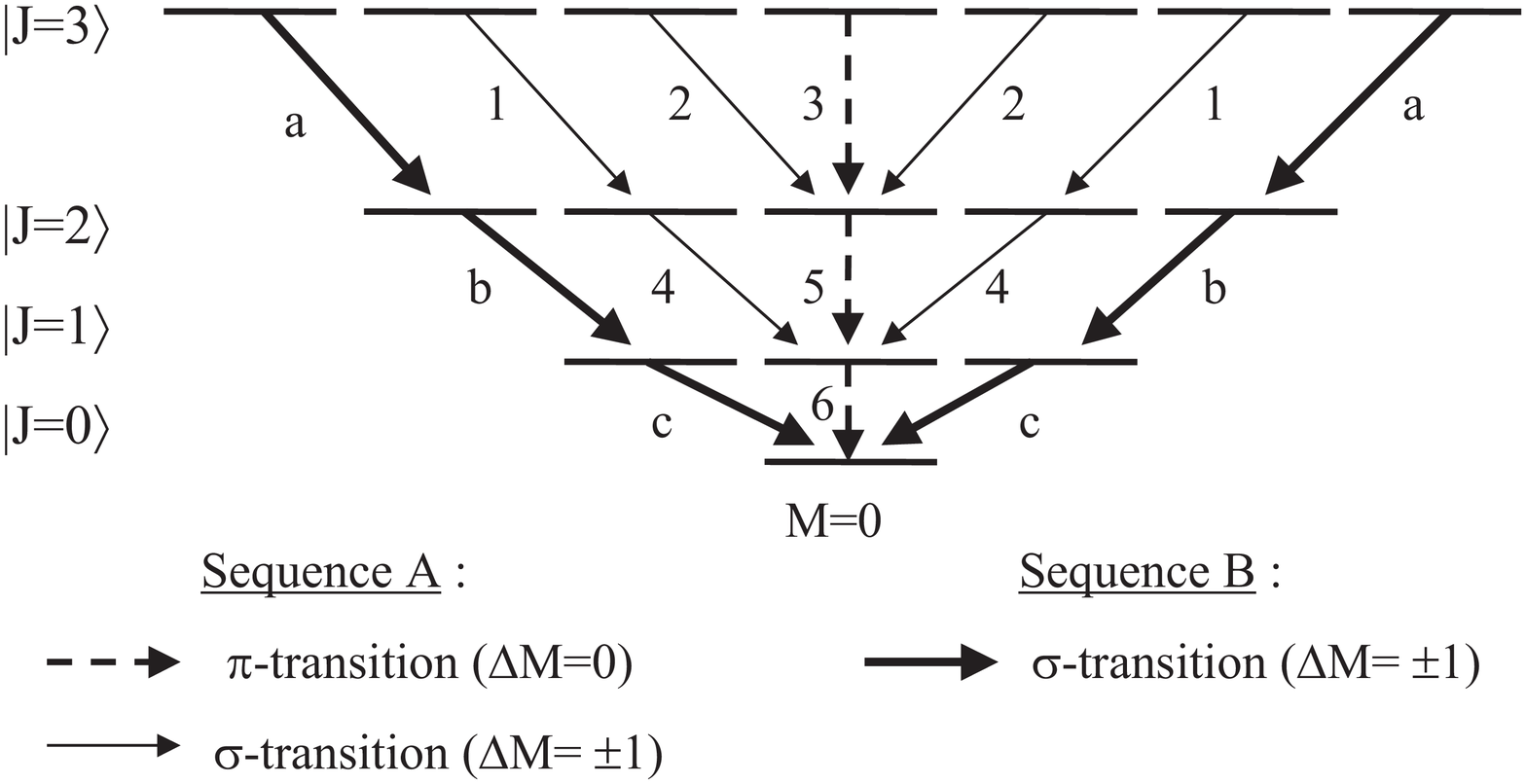} \caption{A
scheme of tuning sequence for rotational cooling using a multiple
field polarization cavity. Sequence A (transition-1 to
transition-6) is used
with the cavity which satisfies $\frac{1}{\protect\lambda _{c}}=(2B_{e}-%
\protect\alpha _{e})$. Sequence B is composed of the 'corner transitions' (a
to c) for cavity with $\frac{1}{\protect\lambda _{c}}=J_{\max }(2B_{e}-%
\protect\alpha _{e})$.}
\label{sequencetune}
\end{figure}

\begin{figure}[t]
\center \epsfxsize=7cm \epsffile{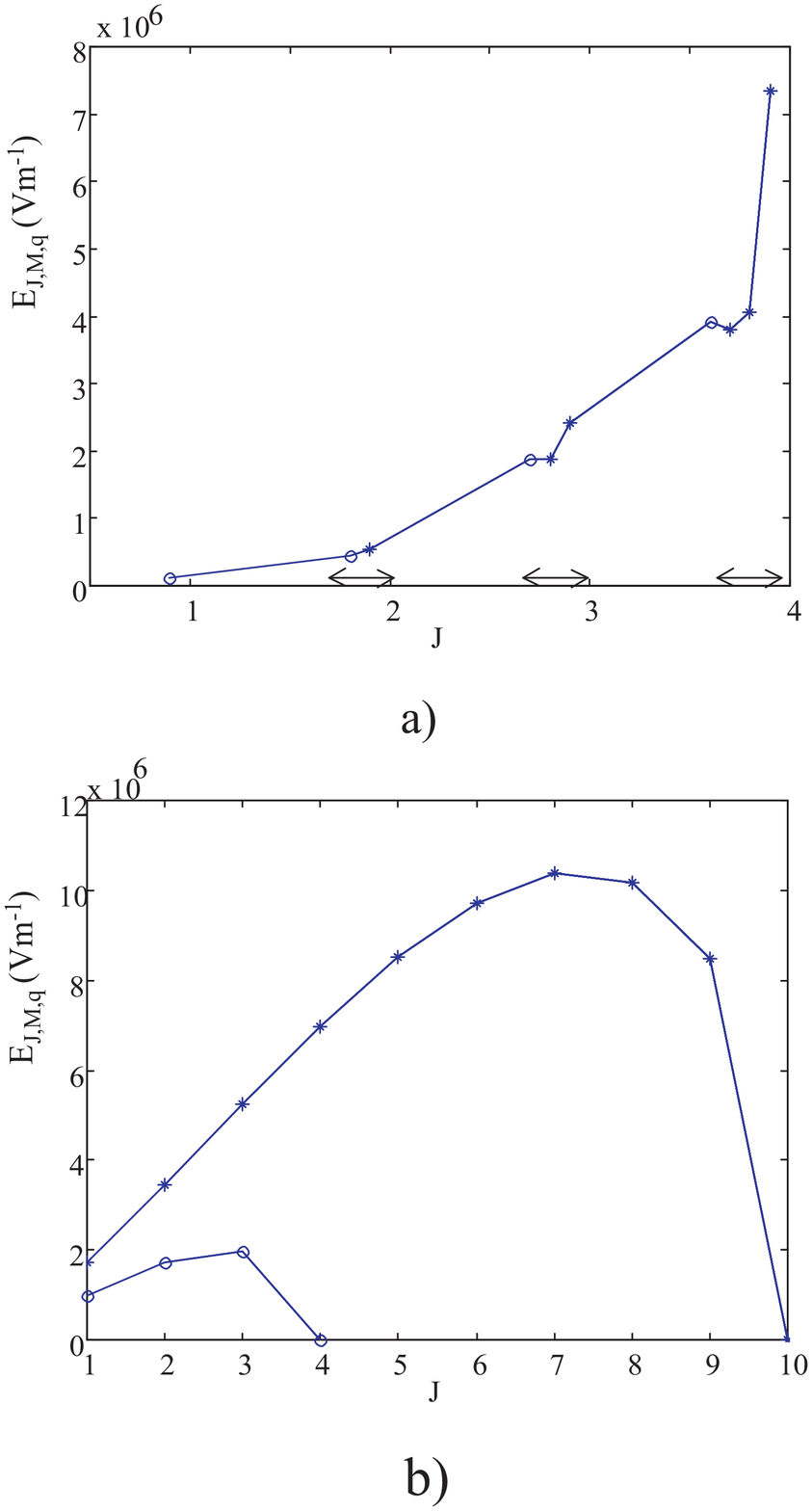} \caption{Electric
fields for tuning the CsF molecules using multiple field
polarization cavity. a)Sequence A - The points corresponding to
each J are displaced slightly from from each other for clarity and
connected by line to show the tuning sequence which starts from
the largest electric field value. The points 'o' correspond to the
$\protect\pi$ transitions. b)Sequence B - The 'o' line is for
$J_{\max }=4$ while the '*' line is for $J_{\max }=10$. The
abscissa J labels the upper rotational state for each transition.}
\label{EJCsFmulti}
\end{figure}

The total number of steps for cooling with multiple polarizations is
generally given by $\allowbreak \frac{1}{2}(J_{\max }-\Omega )(J_{\max
}+\Omega +3)$ for integer $\Omega $ and $\frac{1}{2}(J_{\max }-\Omega
)(J_{\max }+\Omega +2)$ for half-integer $\Omega $ while the total number of
transitions is $\allowbreak J_{\max }(J_{\max }+2)-\Omega (\Omega +2)$ $.$
The possible sequence of tuning is shown in Fig. \ref{sequencetune} for the
molecules with spin $S=0$. The two sequences: A and B correspond to
different cavity parameters given by Eq. \ref{L} and Eq. \ref{L positive}
respectively. Sequence A requires $\allowbreak \frac{1}{2}(J_{\max }-\Omega
)(J_{\max }+\Omega +1)$ steps for integer $\Omega $ and $\frac{1}{2}(J_{\max
}^{2}-\Omega ^{2})$ steps for half-integer $\Omega $ while sequence B
requires $(J_{\max }-\Omega )$ steps. Sequence A cools $\allowbreak J_{\max
}+\Omega +1$ times more states than sequence B but the transitions of both
sequences are complementary. The tuning electric fields for sequence A and
sequence B are calculated from Eq. \ref{E tune final} and Eq. \ref{E tune
positive} as shown in Figs. \ref{EJCsFmulti}a and \ref{EJCsFmulti}b
respectively.

\subsection{Doppler broadening}

The cavity linewidth $\kappa =\omega _{c}/Q$ must be larger than the maximum
Doppler shift in order for the moving molecules to stay in resonance with
the cavity $\omega _{c}/Q>\omega _{D}=\omega _{c}P_{\max }/Mc.$ For thermal
ensemble of OH molecules at about $10K(70ms^{-1})$, the maximum quality
factor we can use is $Q\approx Mc/P_{\max }\approx 10^{6}.$ This value just
gives the highest enhancement factor in the bad cavity regime. So, it is not
helpful to use larger $Q.$ The momentum distribution of the molecules is
essentially unaffected by the small microwave photon recoil momentum from
rotational spontaneous emissions unless the molecules are ultracold.

\subsection{Confinement}

\begin{figure}[t]
\center \epsfxsize=7cm \epsffile{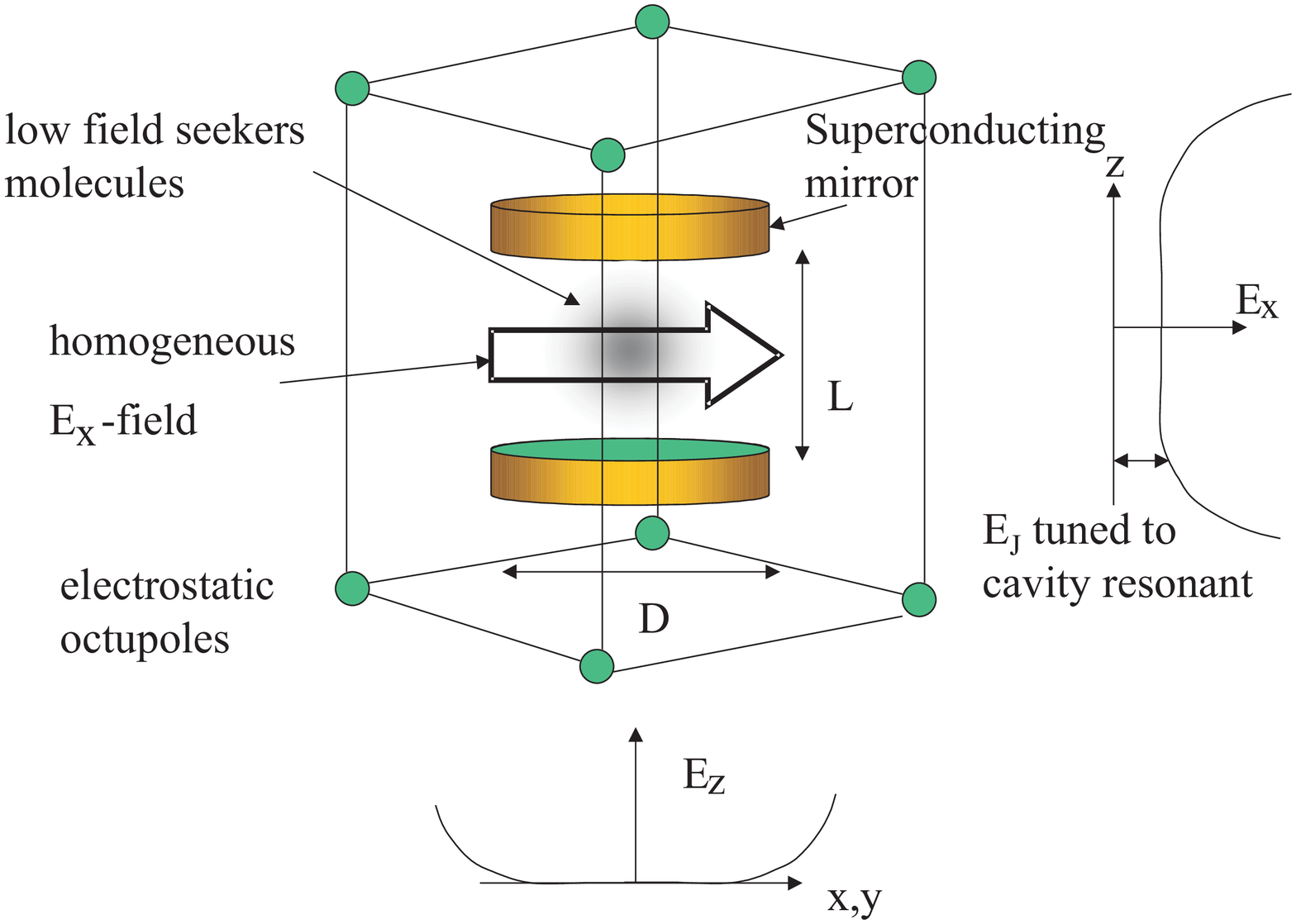}
\caption{Schematic of the apparatus for rotational cooling using
cavity enhanced dissipation and electric field for resonance
tuning and 3-d confinement.} \label{rotcavEtune}
\end{figure}

The maximum diameter (radial dimension) for confocal cavity geometry can be
estimated as $D=\sqrt{3}L$ (see Fig. ~\ref{rotcavEtune})$.$ For a typical
microwave cavity dimension of $1mm$ and molecules with velocity of say $%
70ms^{-1},$ the interaction time of $\Delta t=w/u=0.01ms$\ is much shorter
than the enhanced lifetime. Therefore, we need to confine the molecules
within the cavity.

Figure ~\ref{rotcavEtune} shows the proposed rotational cooling apparatus
with cavity and electric octupoles. Higher order electrostatic poles can be
used to confine the low field seekers with kinetic energy below about 1K
\cite{McAdam}. This background field also prevents the Majorana flop into
untrapping state.

It may be possible to confine molecules using a non-stick solid material
like teflon which has low surface adhesion to certain molecules. However,
this is only applicable to room temperature gas and not translationally cold
gas because collisions with the solid wall will lead to translational
thermalization. Since the molecules in the solid material do not rotate and
have no angular momentum, collisions with the surface do not cause
rotational transition via conservation of angular momentum. The main
advantage of solid material confinement is of course the infinitely long
trapping time.

\subsection{Numerical estimates}

We estimate the cooling time with typical applied tuning field for two types
of molecules only for the $\pi $-transitions involving the middle states ($%
-\Omega \leq M\leq \Omega $). First, we consider $CsF$ \ molecules in state $%
X^{1}\Sigma $ (Fig. \ref{EJCsF}a) with $B_{e}=0.1844cm^{-1}$ and $\alpha
_{e}=1.18\times 10^{-3}cm^{-1},$ and the electric dipole moment $\mu =7.87$%
Debye \cite{Hill Gallagher}. The required electric field is in the order of $%
10^{7}Vm^{-1}$ (Fig. \ref{EJCsF}b). The cavity spacing from Eq. \ref{L} is $%
L=2.04cm$ for mode order $s=1.$ The cavity diameter is $D\sim 3.5cm.$ If the
distance between opposite electrostatic poles which produces the homogeneous
electric field $E_{J}$ is $1cm,$ the required voltage is around $100kV.$ The
free space decay rate is $\Gamma _{o,J,0,0}=\frac{J^{2}}{(2J-1)(2J+1)}%
\allowbreak 9.\,\allowbreak 65\times 10^{-7}s^{-1}$ and for $Q=10^{6},$ the
enhancement factor of $\eta =0.54\times 10^{6}$ boost the decay rate to $%
\Gamma _{c,J,0,0}=0\allowbreak .\,\allowbreak 52\frac{J^{2}}{\left(
2J-1\right) \left( 2J+1\right) }s^{-1}.$ Supposing that a duration of $%
t_{J}=4/\Gamma _{c,J,0,0}\approx 8\frac{\left( 2J-1\right) \left(
2J+1\right) }{J^{2}}$ is allocated for each stage of decay from level $J,$
then the total duration for enchanced decay from $J_{\max }=5$ to $J_{o}=0$
is $t_{tot}=\sum\limits_{J=J_{o}+1}^{J_{\max }}t_{J}=\sum\limits_{J=1}^{5}8%
\frac{\left( 2J-1\right) \left( 2J+1\right) }{J^{2}}\approx 150s$ ($\sim 2.5$
minutes). Since the trapping volume is macroscopically large, many molecules
can be confined and this allows the use of sufficiently dilute gas to reduce
loss rate of molecules due to state changing collisions. Due to the absence
of Majorana flop, a significant number of molecules should remain confined
for long enough time for the rotational cooling to take place. \bigskip

Next, we estimate the essential cooling parameters using OH molecules (with
the levels in Fig. \ref{EJOH}a), which has a large rotational constants $%
B_{e}=18.91cm^{-1}$ and $\alpha _{e}=0.724cm^{-1}$\cite{Smirnov}, and the
static dipole moment of $\mu =1.6676Debye$ \cite{OH dipole moment}. From
Fig. \ref{EJOH}b, we find that the required electric fields are around $%
10^{10}Vm^{-1}.$ Even for cavity dimension of $0.1mm,$ the required voltage
is very large $V\sim 1000kV$ and the free space decay rate is $\Gamma
_{o,J,M,0}=\frac{(J+1/2)(J-1/2)}{(2J-1)(2J+1)}4.\,\allowbreak 452\,1\times
10^{-2}s^{-1}.$ For a small $Q=10^{3}$ and assuming $J_{\max }=5,$ we have $%
\eta =\allowbreak \frac{16\,000}{3\times (3.14)^{2}}=\allowbreak
540.\,\allowbreak 93$ and $t_{tot}=\sum\limits_{J=1}^{5}4\Gamma
_{c,J,M,0}^{-1}=\sum\limits_{J=1}^{5}4\frac{2(2J-1)(2J+1)}{\allowbreak
24(J+1/2)(J-1/2)}\approx 6.\,\allowbreak 7s.$

For OH, the required electric field may be too large to be realized in
practice and it is due to the unusually large rotational constant $B_{e}.$
However, it requires a small cavity Q and the cooling time can be
considerably shortened if the highest Q in the bad cavity regime is used.

\section{Conclusions}

We have proposed a rotational cooling scheme for polar molecules using
Stark-tuned internal levels in a superconducting microwave cavity.
Homogeneous electric field is applied for Stark tuning each transition to
cavity resonance to enhanced rotational spontaneous emissions. Sequential
tuning and the use of the cavity which supports multiple field polarizations
enables all molecules to be cooled towards the ground rotational level.
Numerical estimates for $CsF$ \ molecules show that the scheme requires high
voltage which can be realized using the state-of-the-art technology.
However, molecules with unusually large rotational constants requires
electric field beyond the current capability. The typical cooling time of
one minute with a large Q enhancement in dissipative cavity regime can be
realized for most molecules with moderate rotational constant and large
dipole moment.

\begin{center}
{\large ACKNOWLEDGMENTS}
\end{center}

I gratefully acknowledge support from Deutsche
Forschungsgemeinschaft (Forschergruppe Quantengase) and Optik
Zentrum Konstanz. I thank Dr. Pepijn Pinkse and Prof. Rempe for
hospitality at MPQ where the idea emerged; Priv. Doz. Dr. Peter
Marzlin and Prof. Juergen Audretsch; and Dr. MacAdam for
introducing the Stark ball device. I also thank the referee for
suggesting the extension of the scheme using multiple polarized
cavity.

\newpage

\end{document}